\documentclass[iicol,sn-mathphys-ay]{sn-jnl}

\usepackage{graphicx}%
\usepackage{multirow}%
\usepackage{amsmath,amssymb,amsfonts}%
\usepackage{amsthm}%
\usepackage{mathrsfs}%
\usepackage[title]{appendix}%
\usepackage{xcolor}%
\usepackage{textcomp}%
\usepackage{manyfoot}%
\usepackage{booktabs}%
\usepackage{algorithm}%
\usepackage{algorithmicx}%
\usepackage{algpseudocode}%
\usepackage{listings}%

\usepackage{multicol}
\usepackage{makecell}
\usepackage{multirow}
\usepackage{caption}
\usepackage{bm}
\usepackage[table]{xcolor}

\theoremstyle{thmstyleone}%
\theoremstyle{thmstyletwo}%

\theoremstyle{thmstylethree}%

\begin{document}

\title[Article Title]{Not All Attention is Needed: Parameter and Computation Efficient Tuning for Multi-modal Large Language Models via Effective Attention Skipping}

\def\corrauthemail{zhouyiyi@xmu.edu.cn}


\author[1]{\fnm{Qiong} \sur{Wu}}\email{\textcolor{black}{qiong@stu.xmu.edu.cn}}

\author[*1]{\fnm{Yiyi} \sur{Zhou}}\email{\textcolor{black}{zhouyiyi@xmu.edu.cn}}

\author[1]{\fnm{Weihao} \sur{Ye}}\email{\textcolor{black}{weihaoye@stu.xmu.edu.cn}}

\author[1]{\fnm{Xiaoshuai} \sur{Sun}}\email{\textcolor{black}{xssun@xmu.edu.cn}}

\author[1]{\fnm{Rongrong} \sur{Ji}}\email{\textcolor{black}{rrji@xmu.edu.cn}}

\affil[1]{Key Laboratory of Multimedia Trusted Perception and Efficient Computing, Ministry of Education of China, and also with the Institute of Artificial Intelligence, Xiamen University, 361005, P.R. China.}




\abstract{
Recently, \emph{Multi-modal Large Language Models} (MLLMs) have garnered an influx of interest from both academia and industry.
Despite great progress, MLLMs still have a high demand of downstream task tuning for applications, which consumes excessive parameter and computation overhead. 
In this paper, we propose a novel parameter and computation efficient tuning method for MLLMs, termed \emph{Effective Attention Skipping} (EAS). 
Concretely, we first reveal that \emph{multi-head attentions} (MHAs) in MLLMs, the primary source of computation, are often redundant to downstream tasks. 
Based on this observation, EAS evaluates attention redundancy and skips the less important MHAs to speed up inference.
Besides, we also propose a novel \emph{propagation-of-information adapter} (PIA) to serve attention skipping while maintaining parameter efficiency. 
More importantly, PIA can be further re-parameterized into \emph{feed-forward networks} (FFNs) for zero-extra latency. 
To validate EAS, we apply it to three common MLLMs, and conduct extensive experiments on six downstream tasks.
The experimental results show that EAS can not only retain the high performance of MLLMs but also reduce the updated parameters scale greatly, while speeding up inference speed to a large extent. 
For instance, LLaVA-EAS can obtain 96.2\% accuracy while accelerating the inference speed by about +20\% on ScienceQA.
Our code is publicly released at \url{https://github.com/DoubtedSteam/EAS}.
}




\maketitle

\section{Introduction}\label{sec1}

Recently, the great success of \emph{Large Language Models} (LLMs) also sparks an influx of interest in extending these giant models to more modalities, \emph{i.e.}, \emph{Multi-modal Large Language Models} (MLLMs) \cite{luo2023cheap, liu2023visual, li2023blip, llava-1.5, Internvl}.
However, compared with the unimodal LLMs, MLLMs inevitably require much more computation with the introduction of new modalities \cite{bao2022vlmo, alayrac2022flamingo, li2023blip, liu2023visual}.
For instance, compared with LLaMA \cite{touvron2023llama}, LLaVA \cite{llava-1.5} takes about 6.2 times the computation overhead to solve the questions from the ScienceQA \cite{lu2022learn}.
Moreover, the scale of available data for MLLMs is much smaller than LLMs \cite{zhang2023llama, gao2023llama, wu2023parameter}, especially the manually labeled SFT examples.
In this case, the generalization of MLLMs is often limited, and the requirement of downstream task adaption is often demanded. 

Recently, numerous efforts are devoted to the efficient tuning of MLLMs \cite{conf/iclr/HeZMBN22, conf/cvpr/Sung0B22, conf/iclr/HuSWALWWC22}.
Among these advancements, \cite{wu2023parameter} propose a non-trivial task for large-scaled pre-trained models, termed \emph{parameter and computation efficient transfer learning} (PCETL). 
A key intuition behind PCETL is that the sheer size of MLLMs' parameters and computation is critical for large-scale pre-training but redundant to specific tasks. 
In this case, for downstream task adaptions, PCETL not only requires to reduce the number of updated parameters, akin to PETL \cite{houlsby2019parameter, hu2022lora, sung2022vl, journals/ijcv/ZhaoWZ00S24}, but also needs to remove the redundant modules for better efficiency. 

\begin{figure}[t]
\centering
\includegraphics[width=\linewidth]{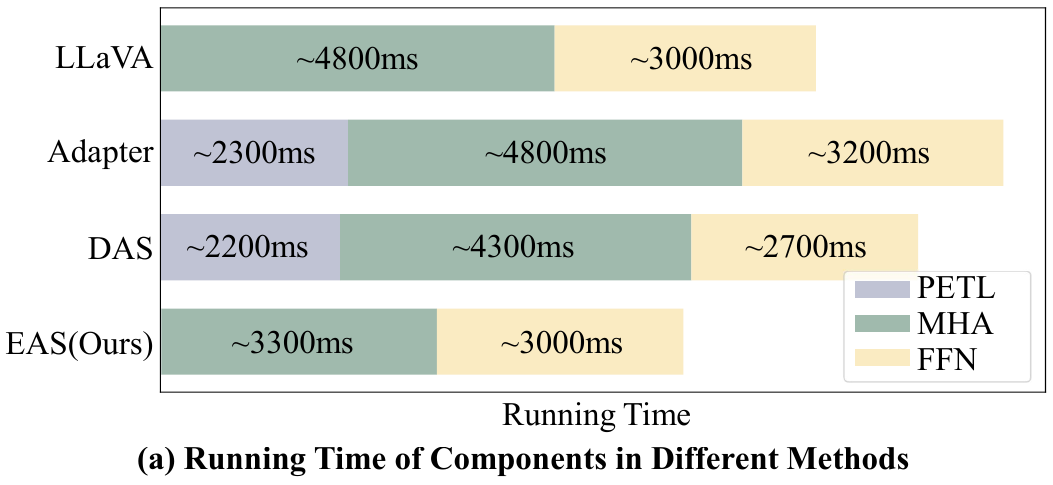}
\label{fig:1a}
\includegraphics[width=\linewidth]{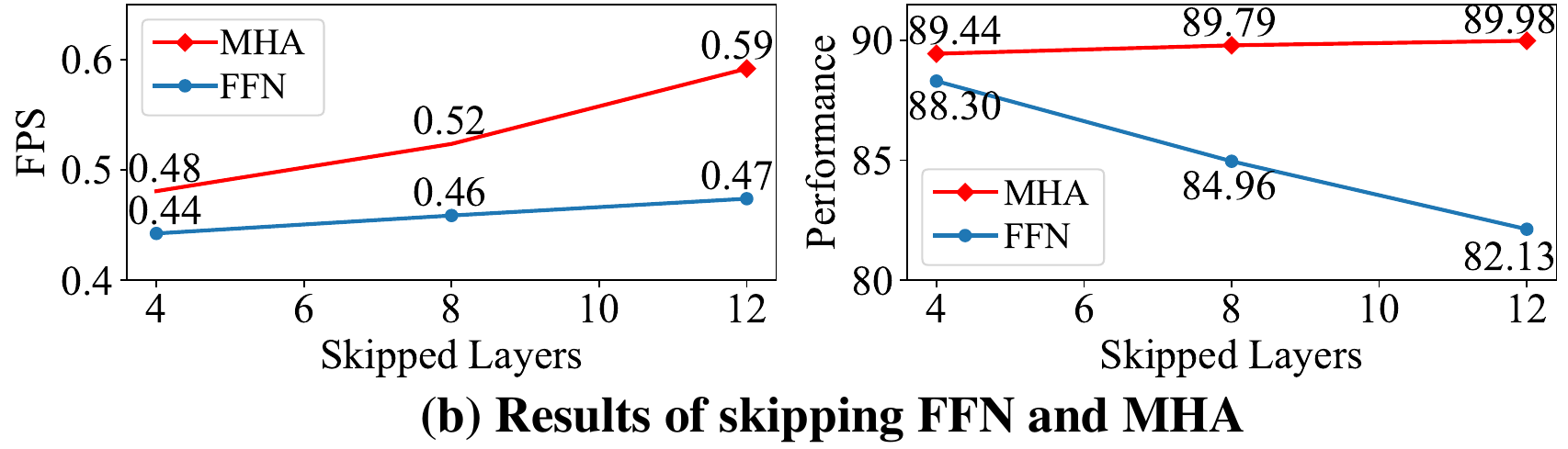}
\label{fig:1b}
\caption{
(a) Latency comparison between EAS and other PETL methods, including Adapter \cite{sung2022vl} and DAS \cite{wu2023parameter}, and the default LLaVA \cite{conf/nips/LiuLWL23a}. 
(b) Performance and speed comparisons of skipping different numbers of MHA and FFN by our EAS on ScienceQA.
}
\label{fig2:TimeCost}
\end{figure}

\begin{figure}[t]
\centering
\includegraphics[width=\linewidth]{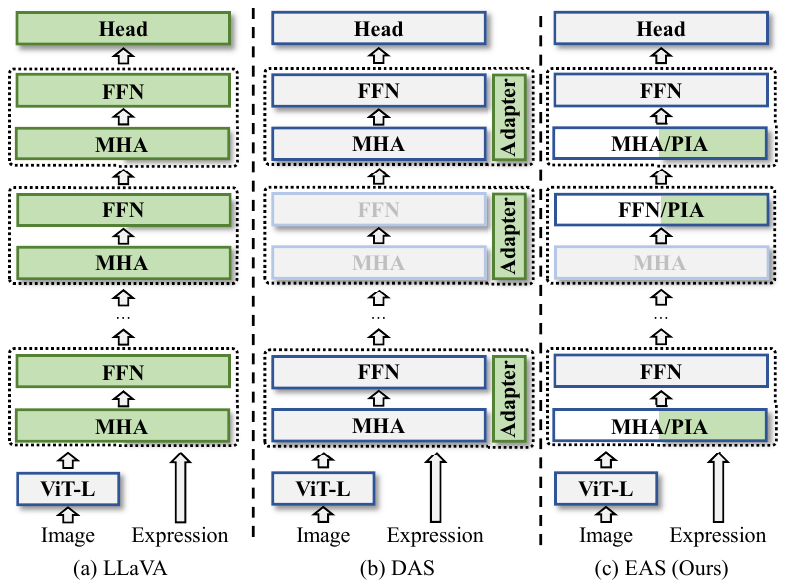}
\caption{
Illustrations of LaVIN, DAS and our EAS. 
(a) LaVIN inserts lightweight adapters before MHAs for multi-modal adaption.
(b) DAS skips redundant Transformer layers of LaVIN, but still incurs extra latency.
(c) EAS resort to skipping MHAs, to achieve true model acceleration with the proposed PIA.
}
\label{fig1:Motivation}
\end{figure}

However, achieving the target of PCETL still remains challenging.
On the one hand, the redundancy of MLLMs' components for downstream tasks still requires more in-depth exploration, \emph{e.g.}, \emph{multi-head attention} (MHA) and \emph{feed-forward network} (FFN). 
As noted in existing literature \cite{martins2020sparse, tolstikhin2021mlp}, MHA mainly acts a role of dependency modeling, while FFN is designed to enhance model capacity.
Furthermore, MHA consumes 35\% more time than FFN despite having only half the number of parameters, as shown in Fig.~\ref{fig2:TimeCost}-(a).
In \cite{wu2023parameter}, Wu \emph{et al.} take the entire transformer layers as a candidate to evaluate redundancy.
In this coarse-grained way, only a few candidates can be skipped, limiting the merits of PCETL.
Meanwhile, we also observe that skipping a certain number of MHAs does not affect performance, while it will result in a drastic drop when skipping FFNs, as shown in Fig.~\ref{fig2:TimeCost}-(b).

On the other hand, how to acheive true speed-up of MLLMs remains an open problem for PCETL. 
Specifically, although the lightweight adapters used to replace the original module  only consume a few computational overhead \cite{wu2023parameter}, but the incurred  latency is non-negligible.
As revealed in recent progresses \cite{luo2023cheap, gao2023llama}, the additional branches added by adapters inevitably slow down inference although they only brings a few of FLOPs \cite{luo2023cheap}.
The same as shown in Fig.\ref{fig2:TimeCost}, the latency caused by adapters takes up to 22.3\% inference time of Adapter \cite{sung2022vl}.
Conclusively, correctly judging the redundancy of MLLMs to achieve true speed-up is one main obstacle of PCETL.

To handle these challenges, we propose a novel approach for the parameter and computation efficient tuning of MLLMs in this paper, termed \emph{Effective Attention Skipping} (EAS). 
In particular, we first investigate the redundancy of different components of MLLMs, and show that skipping MHA can reduce more computation overhead without impeding performance, as shown in Fig. \ref{fig2:TimeCost}-(b).
The principle of EAS is to only skip the redundant MHAs for better efficiency and performance. 
To avoid the latency caused by adapters \cite{wu2023parameter}, we also equip EAS with an innovative parameter efficient module called \emph{propagation-of-information adapter} (PIA), which can not only replace the skipped MHAs for adaption, but also can be seamlessly re-parameterized into FFNs for zero extra cost during inference. 
With these innovative designs, EAS can help MLLMs better approach the targets of PCETL on vision-language tasks, \emph{i.e.}, true model acceleration with a small number of trainable parameters. 

To validate EAS, we first apply it to two MLLMs, \emph{i.e.}, LLaVA \cite{llava-1.5} and LaVIN \cite{luo2023cheap}, and conduct extensive experiments on two out-domain benchmarks, Slake and AID, and the open-set multi-modal question answer benchmark, namely ScienceQA \cite{lu2022learn}.
To align DAS \cite{wu2023parameter}, we also generalize EAS to a representative VL pre-trained model called METER \cite{conf/cvpr/DouXGWWWZZYP0022} on three VL benchmarks, \emph{i.e.}, VQA2.0 \cite{conf/cvpr/GoyalKSBP17}, NLVR$^2$ \cite{conf/acl/SuhrZZZBA19} and Flickr30K \cite{journals/ijcv/PlummerWCCHL17}.
The experimental results show that EAS can not only retain high performance and parameter efficiency against existing PETL and PCETL methods \cite{zhang2023llama, luo2023cheap, wu2023parameter}, but also significantly speed up inference speed on downstream tasks.
For instance, EAS can improve the inference speed by $1.20\times$ to the default LLaVA \cite{liu2023visual} without performance degradations.

Overall, our contributions are three-fold:
\begin{itemize}
\item We propose a novel parameter and computation efficient tuning method for MLLMs, termed \emph{Effective Attention Skipping} (EAS), which retains the high performance of MLLMs and reduces both parameter and computation expenditures on downstream tasks.

\item We propose a novel \emph{propagation-of-information adapter} (PIA) that can be used to serve attention skipping and be fully re-parameterized into MLLMs for true model acceleration.  

\item We apply our EAS to two recent MLLM and a representative VLP model, \emph{i.e.}, LLaVA, LaVIN and METER, on six VL benchmarks. The experiments show the obvious merits of EAS in both parameter and computation efficiencies, \emph{e.g.}, speeding up LaVIN by 2.18$\times$ without performance degradations. 

\end{itemize}

\begin{figure*}[t]
\centering
\includegraphics[width=1.0\textwidth]{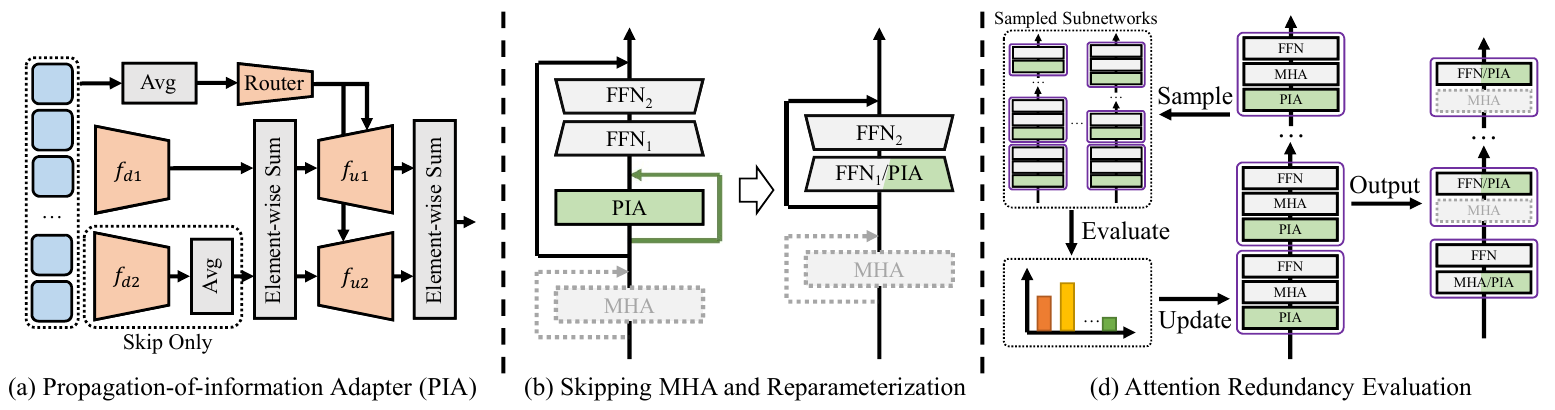}
\caption{
 Illustrations of the proposed \emph{Effective Attention Skipping} (EAS).
(a) The architecture of \emph{propagation-of-information adapter} (PIA).
PIA uses a multi-path design for up- and down-samplings, which can act a role of MHA in information exchange for MLLMs.   
(b) The deployment of PIA. 
PIA can serve to replace the skipped MHA as a parameter efficient method for task adaption. 
After training, its parameters can be re-parameterized into FFN, incurring no extra latency. 
(c) The process of attention redundancy evaluation. 
EAS also adopts a $k$-armed bandit based algorithm for the automatic redundancy evaluation on MHAs of MLLMs. After evaluation, we skip the redundant MHAs with PIAs.
}
\label{Fig3:Framework}
\end{figure*}

\section{Related Work}

\subsection{Vision-language pre-trained  models}

Recent years have witnessed the rapid development of \emph{vision-language pre-trained} (VLP) models \cite{li2019visualbert, lu2019vilbert, su2019vl, cho2021unifying, conf/icml/WangJLZCYL22} on various vision-language tasks \cite{journals/ijcv/GoyalKASBP19, lu2022learn, journals/ijcv/JiangLZ24}.
To facilitate vision-language alignment and learn generalized multi-modal representations, VLP models also adopt self-supervised objects and conduct pre-training on massive image-text pairs, such as \emph{masked language modeling} \cite{devlin2018bert, liu2019roberta}, \emph{masked image modeling} \cite{conf/iclr/Bao0PW22, conf/icml/ChenRC0JLS20, conf/iclr/DosovitskiyB0WZ21} and~\emph{image-text matching} \cite{conf/cvpr/DouXGWWWZZYP0022, conf/icml/KimSK21}.
With the advent of \emph{large language models} (LLMs) \cite{zhang2205opt, touvron2023llama, openai2023gpt}, recent advance \cite{luo2023cheap,liu2023visual,li2023blip, dong2021dual} focus on extending these LLMs to vision-language tasks \cite{guo2024benchmarking, marino2019ok, conf/cvpr/GoyalKSBP17}, \emph{i.e.}, \emph{multi-modal large language models} (MLLMs).
For instance, BLIP-2 \cite{conf/icml/0008LSH23} introduces a QFormer, to bridge the gap between vision and language modalities, selecting relevant visual semantics as additional tokens for LLMs.
MINI-GPT4 \cite{journals/corr/abs-2304-10592} uses a projection layer to map visual features into the language model.
LLaVA \cite{conf/nips/LiuLWL23a} also follows this principle, but proposes a carefully designed training scheme in addition.
Meanwhile, some efforts are also devoted to parameter-efficiently transfer learning of LLMs to vision-language tasks \cite{goyal2017making, lu2022learn}. 
These methods insert lightweight modules like adapter \cite{zhang2023llama} for downstream adaption instead of fully tuning LLMs \cite{hu2022lora}.
Meanwhile, the use of visual tokens greatly exacerbates the already high computation of MLLMs \cite{liu2023visual, zhang2023llama}.
In this paper, we focus on both parameter and computation efficient tuning for MLLMs.  

\subsection{Parameter and Computation Efficient Tuning}

\emph{Parameter-efficient transfer learning} (PETL) is proposed to save the training and storage costs of LLMs \cite{brown2020language, li2021prefix, hu2022lora}.
The principle of PETL is to transfer the pre-trained models to downstream tasks with only a small number of trainable parameters \cite{brown2020language, li2021prefix}, and its methodologies can be roughly divided into three main categories, \emph{i.e.} \emph{prompt tuning} \cite{jia2022visual, zhou2022learning, journals/ijcv/BulatT24}, \emph{adapter} \cite{karimi2021compacter, gao2023clip} and the \emph{re-parameterized methods} \cite{hu2022lora, luo2023towardsefficient}.
These methods mainly focus on efficiently transferring models to the downstream tasks.
PEQA \cite{kim2023memory} proposes a simple method that only updates quantization scales with quantized LLMs to reduce the requirements of memory during the training.
QLoRA \cite{dettmers2023qlora} quantizes and freezes the original model, and uses LoRA to transfer the model to downstream tasks.
HydraLoRA \cite{tian2024hydralora} uses an asymmetric structure based on LoRA to reduce the need for domain expertise in transferring.
Recently, \cite{wu2023parameter} propose a new learning task called \emph{parameter and computation efficient transfer learning} (PCETL), which further requires to reduce model redundancy based on the setting of PETL.
\cite{wu2023parameter} also propose a strong baseline called \emph{Dynamic Architecture Skipping} (DAS) for PCETL, which evaluates the redundancy of Transformer layers by a reinforcement learning method and replaces the most redundant layers with adapters \cite{sung2022vl}.
Motivated by \cite{hu2022lora, wu2023parameter}, we also focus on PCETL for MLLMs in this paper. 
We propose a granular exploration for MLLMs, and address the latency caused by using adapters for skip connections.

\subsection{Improving Efficiency in MLLMs}
The inference efficiency of the MLLM has also attracted much attention from researchers. 
In recent years, many works focus on improving the inference efficiency of MLLM, with the main methods including simplifying the structure of models \cite{tong2024flashsloth, zhou2024tinyllava, zhu2024llava} and reducing the length of the inference sequence \cite{zhang2024cls, han2024rethinking}.
In the way to reduce the inference sequence, FastV~\cite{chen2024image} uses a metric to evaluate the importance of visual tokens and prunes half of them to reduce computation overhead.
FitPrune~\cite{ye2025fit} removes visual tokens without affecting performance by fitting changes in the attention matrix.
However, these methods mainly focus on reducing the number of visual tokens, and efficiency improvement is mainly in the prefilling stage.
In terms of structural design, recent MLLMs like MoE-LLaVA \cite{moe_llava} selectively activate expert parameters through the MoE structure.
Mono-InternVL~\cite{luo2024mono} splits the parameters into text and vision experts to improve the efficiency. 
And RoE~\cite{wu2024routing} sparses the model through selectively activating the layers in inference.
However, these methods rely on fully tuning MLLMs. 
In this paper, we achieve both parameter and computation efficiency in transferring MLLMs to the downstream task.  

\section{Preliminary}

We first recap the principle of \emph{parameter and computation efficient transfer learning} (PCETL) \cite{wu2023parameter} for \emph{multi-modal large language models} (MLLMs). 
Concretely, given an MLLM $G(\cdot|\theta)$, which consists $n$ Transformer layers, and its whole parameters denoted as $\mathbf{\theta}=\{\theta_1, \theta_2, ..., \theta_n\}$, the objective of PCETL can be defined by 
\begin{equation}
\begin{aligned}
\operatorname*{argmin}_{\bm{\sigma}, \mathcal{K}} \mathcal{L}\big(G(I,T|\bm{\theta}_\mathcal{K}, \bm{\sigma})\big),
\end{aligned}
\label{PCETL_principle}
\end{equation}
where $\bm{\theta}_\mathcal{K} = \{\theta_{k_1}, \theta_{k_2}, ..., \theta_{k_m}\} \in \bm{\theta}$ are the parameters of a sub-network of MLLM and $\sigma$ is a small number of parameters for model tuning.

From Eq.\ref{PCETL_principle}, we can see that PCETL needs to reduce parameter cost during adaption, just like PETL \cite{li2021prefix, hu2022lora, sung2022vl}.
Meanwhile, it also requires practitioners to speed up inference by evaluating and removing the redundant modules.
Compared with previous model compression and acceleration tasks \cite{krishnamoorthi2018quantizing, meng2020pruning,  dong2021attention},  one difference of PCETL is that the original model weights should keep unchanged, so as to facilitate more task adaptions, which is essential for the giant models that are expensively pre-trained with massive data. 

\cite{wu2023parameter} also propose a strong baseline called \emph{dynamic architecture skipping} (DAS), which considers the entire Transformer layer for redundancy evaluation.
As discussed above, the components in MLLMs often serve different roles, \emph{i.e.}, MHA and FFN. 
In this case, the granular evaluations are beneficial for PCETL.
Besides, the skip connections in DAS are aided by adapters \cite{houlsby2019parameter, sung2022vl} which also incur non-negligible latency during the inference \cite{wu2023parameter, hu2022lora}.

\section{Effective Attention Skipping}

\subsection{Overview}

To achieve parameter and computation efficient tuning for MLLMs, we propose an \emph{Effective Attention Skipping} (EAS) approach in this paper, as illustrated in Fig.~\ref{Fig3:Framework}.

Concretely, given an MLLMs $G(\cdot|\theta)$, EAS aims to remove its redundant components for downstream task adaption. Different to DAS \cite{wu2023parameter}, the evaluation target of EAS is more specific, \emph{i.e.}, MHAs of a MLLM. 
Thus, the objective can be defined by 
\begin{equation}
\begin{aligned}
\operatorname*{argmin}_{\mathcal{K}} \mathcal{L}\big(G(I,T|\bm{\theta}^{A}_\mathcal{K}, \bm{\theta}^{F})\big),
\end{aligned}
\end{equation}
where $\bm{\theta}^{A}_\mathcal{K}$ represent a subset of MHA modules and $\bm{\theta}^{F}$ denote all FFN modules.
In practice, similar to DAS \cite{wu2023parameter}, we first conduct a reinforcement learning based redundancy evaluation on MHAs for downstream tasks.
Afterwards, given the redundancy scores, we determine which MHAs to skip, and replace them with an adapter-based connection.

However, adapters will account for excessive inference time, as shown in Fig.~\ref{fig2:TimeCost}-(a).
The widely used re-parameterized methods like LoRA \cite{hu2022lora} can approximate the QKV projections but cannot replace the entire MHA for skip connections \cite{luo2023towardslanguage}.

To this end, we propose a novel \emph{Propagation-of-Information Adapter} (PIA) to achieve zero-cost skip connections, which can propagate averaged global features for information exchange, and be reparameterized into FFNs during inference.

In this case, the objective of EAS can be formulated as
\begin{equation}
\begin{aligned}
\operatorname*{argmin}_{\bm{\sigma}, \mathcal{K}} \mathcal{L}\big(G(I,T|\bm{\theta}^{A}_\mathcal{K}, \bm{\theta}^{F} + \bm{\sigma})\big),
\end{aligned}
\end{equation}
where, $\bm{\sigma}$ is the reparameterizable parameters of PIA.
With PIA, we can adapt MLLM to downstream tasks while speeding up inference. 

\subsection{Propagation-of-Information Adapter}

When redundant MHAs are identified, the next key step is to effectively skip them without additional latency. 
To approach this target, we propose a novel re-paremeterizable adapter, termed \emph{Propagation-of-Information Adapter} (PIA). 

As shown in Fig. \ref{Fig3:Framework}-(a), PIA also adopts a bottleneck structure to scale down the hidden dimension of the inputs, akin to existing adapters \cite{sung2022vl, luo2023towardsefficient}, thereby achieving a low-rank approximation of full tuning \cite{hu2022lora, wu2023approximated}. 
However, it also differs in its inner path designs. 

Concretely, given the input features of the $i^{th}$ layer, denoted as $\mathbf{X}^{(i)} \in \mathbb{R}^{n \times d}$,
PIA first projects $\mathbf{X}^{(i)}$ onto two separate low-dimensional semantic spaces, and obtain the hidden features  $\mathbf{H}^{(i)}$ of PIA by
\begin{equation}
\begin{aligned}
\mathbf{H}^{(i)} = f_{d1}(\mathbf{X}^{(i)}) + avg(f_{d2}(\mathbf{X}^{(i)})),
\end{aligned}
\label{first_half_PIA}
\end{equation}
where $f_{d1}(\cdot)$ and $f_{d2}(\cdot)$ are the two linear projections, and $avg(\cdot)$ refers to average pooling. 
$\mathbf{H}^{(i)}\in \mathbb{R}^{n \times r}$ has a much smaller dimension than $\mathbf{X}^{(i)}$, \emph{i.e.,} $r \ll d$.
Besides, with the information propagation through the average pooling, the skipping of MHA can lead to less impact.

In addition to the low-dimension projection, Eq.~\ref{first_half_PIA} also realizes information exchange among all input tokens via the combination with the averaged feature. 

During up-sampling, we also use two separate linear projections for the hidden features, and adopt a path router for their weighted combination. 
Thus, the output $\mathbf{X}^{(i)'}$ is obtained by
\begin{equation}
\begin{aligned}
\mathbf{X}^{(i)'} &= \alpha_1 f_{u1}(\mathbf{H}^{(i)}) + \alpha_2 f_{u2}(\mathbf{H}^{(i)}), \\
\mathrm{where} & \ \ [\alpha_1, \alpha_2] =\mathrm{router}(avg(\mathbf{X}^{(i)})).
\end{aligned}    
\label{last_half_PIA}
\end{equation}
Here, $f_{u1}(\cdot)$ and $f_{u2}(\cdot)$ are two linear projections.
Their weights will be given by the router in the prefilling stage.
And $\mathrm{router}(\cdot)$ is the routing function defined by
\begin{equation}
\bm{\alpha} = softmax(\frac{\hat{\mathbf{x}}^{(i)}\mathbf{W}_r + \mathbf{b}_r}{\tau}),
\end{equation}
where $\hat{\mathbf{x}}^{(i)} = avg(\mathbf{X}^{(i)})$, $\mathbf{W}_r \in \mathbb{R}^{d \times 2}$ is the weight matrix, $\tau$ is the temperature of \emph{softmax}.

Overall, the multi-path design can still hold parameter efficiency via low-dimensional projections, acting the role of MHA in information exchange \cite{vaswani2017attention}.

\textbf{Re-parameterization.} With the effective structure, PIA also needs to re-parameterize its weights into the models to further improve computation efficiency. 
Here, we embed PIA into FFN, and avoid the skip connections \cite{wu2023parameter} to retain network complexity. 

Concretely, we first place PIA in the residual connection of FFN, as shown in Fig. \ref{Fig3:Framework}-(b):
\begin{equation}
\begin{aligned}
\mathbf{X}^{(i+1)} = \mathbf{X}^{(i)} + \mathrm{FFN}\big(\mathbf{X}^{(i)} + \mathrm{PIA}(\mathbf{X}^{(i)})\big).
\end{aligned}
\end{equation}
To re-parameterize PIA into the nearby projection weights of FFN \cite{hu2022lora,luo2022towards}, we still need to convert PIA into one linear layer:
\begin{equation}
\begin{aligned}
\mathrm{PIA}(\mathbf{X}^{(i)})=\mathbf{X}^{(i)}\mathbf{W}_p + \mathbf{b}_p.
\end{aligned}
\end{equation}

More specifically, we aim to simplify the definition of PIA in Eq. \ref{first_half_PIA} and \ref{last_half_PIA} by
\begin{equation}
\mathbf{X}^{(i)'}=(\mathbf{X}^{(i)}\mathbf{W}_d+\mathbf{b}_d)\mathbf{W}_u+\mathbf{b}_u,
\end{equation}
where $\mathbf{W}_d$ and $\mathbf{W}_u$ are new projection matrices for down- and up-sampling, and $\mathbf{b}_d$, $\mathbf{b}_u$ are biases.   

In particular, the down-sampling for the hidden state $\mathbf{H}^{(i)}$ is expected to be
\begin{equation}
\begin{aligned}
\mathbf{H}^{(i)} =  \mathbf{X}^{(i)}\mathbf{W}_{d} + \mathbf{b}_{d}.
\end{aligned}
\end{equation}
However, in Eq.\ref{first_half_PIA}, the down-sampling uses a two-path design for information exchange. 
In this case, we obtain $\mathbf{W}_d$ and $\mathbf{b}_d$ via 
\begin{equation}
\begin{aligned}
\mathbf{W}_{d} &= \mathbf{W}_{d1}, \\
\mathbf{b}_{d} &= \mathbf{b}_{d1} + avg(f_{d2}(\mathbf{X}^{(i)})),
\end{aligned}
\label{reparameter_down}
\end{equation}
where $\mathbf{W}_{d1} \in \mathbb{R}^{d \times r}$ and $\mathbf{b}_{d1} \in \mathbb{R}^{r}$ are the projection matrix and bias vector of $f_{d1}(\cdot)$ in Eq.~\ref{first_half_PIA}.
Via Eq.\ref{reparameter_down}, we can merge the two-path design in one linear projection. 

In practice, $\mathbf{b}_d$ is frozen after the first decoding step of MLLMs regardless of the change of $\mathbf{X}^{(i)}$.
This design can enable PIA to achieve information exchange during inference without hindering re-parameterization. 

Similar to down-sampling, up-sampling in Eq.\ref{last_half_PIA}, is redefined by
\begin{equation}
\begin{aligned}
\mathbf{X}^{(i)'}=& \mathbf{H}^{(i)}\mathbf{W}_{u} + \mathbf{b}_{u}, \\
\mathrm{where} \ \mathbf{W}_{u}=& \alpha_1\mathbf{W}_{u1} + \alpha_2\mathbf{W}_{u2}, \\
                 \mathbf{b}_{u}=& \alpha_1\mathbf{b}_{u1} + \alpha_2\mathbf{b}_{u2}.
\end{aligned}
\label{reparameter_up}
\end{equation}
Here, $\alpha_1$ and $\alpha_2$ are the routing weights in Eq. \ref{last_half_PIA}.
$\mathbf{W}_{u1} \in \mathbb{R}^{r \times d}$, $\mathbf{W}_{u2} \in \mathbb{R}^{r \times d}$ are projection matrices  of $f_{u1}(\cdot)$ and $f_{u2}(\cdot)$ in Eq.\ref{last_half_PIA}, and $\mathbf{b}_{u1}, \mathbf{b}_{u2} \in \mathbb{R}^{d}$ are bias vectors.

After obtaining the merged $\mathbf{W}_d$ and $\mathbf{W}_u$, we can then transform PIA into one linear layer via
\begin{equation}
\begin{aligned}
\mathbf{X}^{(i)'} =& (\mathbf{X}^{(i)}\mathbf{W}_{d} + \mathbf{b}_{d})\mathbf{W}_{u} + \mathbf{b}_{u} \\
              =& \mathbf{X}^{(i)}\mathbf{W}_{d}\mathbf{W}_{u} + \mathbf{b}_{d}\mathbf{W}_{u} + \mathbf{b}_{u} \\
              =& \mathbf{X}^{(i)}\mathbf{W}_{p} + \mathbf{b}_{p}.
\end{aligned}
\label{reparameter_overall}
\end{equation}
Here, $\mathbf{W}_p$ is the approximated low-rank weight matrix, and it then can be re-parameterized into the nearby weight matrices of FFN, like \cite{hu2022lora, luo2023towardsefficient}.

Note that, the average feature in Eq.~\ref{first_half_PIA}, \emph{i.e.} $\hat{\mathbf{x}}^{(i)} = avg(\mathbf{X}^{(i)})$, and the path router are example-dependent. 
In this case, the re-parameterization of PIA will be executed after the first decoding step of MLLMs. 
Considering the fact that most MLLMs often needs to decode a long sequence \cite{luo2023cheap, zhang2023llama}, PIA can still save massive computation during inference.    

\subsection{Attention Redundancy Evaluation}

In EAS, we adopt a $k$-arm bandit based algorithm to automatically evaluate the redundancy of MHAs in MLLMs for downstream task.
This formulation addresses the combinatorial challenge of selecting $k$ redundant modules from $n$ MHAs.
The search space of $\binom{n}{c}$ subnetworks is computationally hard to evaluate exhaustively.
The k-armed bandit framework efficiently balances exploration of novel configurations with exploitation of high-performing subnetworks while providing regret-bounded optimization guarantees \cite{auer2002finite}.
In this way, we can obtain the accelerated model with the best performance at a given computational overhead.

Concretely, given a MLLM, we first randomly sample and tune the subnetworks with PIAs as adapters for a few epochs.
Afterwards, we initialize a numerical action preference, denoted as $\mathbf{s} \in \mathbb{R}^n$, and then we keep training and testing the MLLM with PIAs. 
Every few steps, we will sample several substructures for comparison and update the action preference.
In the comparison, we skip $k$ MHAs according to $\mathbf{s}$, and the action policy at the $t$ step is obtained by
\begin{equation}
\begin{aligned}
\pi^{(t)}_i \sim U(0, \frac{e^{s^{(t-1)}_i}}{\sum_k e^{s^{(t-1)}_k}}),
\label{get_score}
\end{aligned}
\end{equation}
where $U(a,b)$ is the uniform distribution between $a$ and $b$.
In this way, the effective modules are more likely to obtain a higher preference score.
The action preference score $s_i^{(t)}$ directly quantifies the importance of the $i$-th MHA module.
Lower scores indicate higher redundancy.
For each subnetwork, we skip $k$ MHA modules with the lowest action policies:
\begin{equation}
\begin{aligned}
& \rho^{(t)} = \operatorname*{argmin}_{\{\mu_1, \mu_2, ..., \mu_k\}} \sum_{i \in \{\mu_1, \mu_2, ..., \mu_k\}}\pi^{(t)}_{i}, \\
& \mathrm{where} \ |\rho| = k, 1 \leq \mu_j \leq n.
\label{get_arch}
\end{aligned}
\end{equation}
Here, $\rho^{(t)}$ is the index of the skipped MHA modules.
The MHA with a greater $\mathbf{s}_i^{(t)}$ will be kept.

Then, we update the action preference of each MHA through comparisons.
For each $T$ step, 
we first sample the action policies for these $m$ subnetworks, \emph{e.g.}, the action policy for the $j$-th subnetwork $\mathbf{\Pi}^{(t)}_j=[\pi^{(t)}_{j,1}, \pi^{(t)}_{j,2}, ..., \pi^{(t)}_{j,n}]$, according to Eq.\ref{get_score}.
Based on this, we can gain the skipped MHAs in these $m$ subnetworks,
\emph{i.e}, $\Phi^{(t)} = \{\rho^{(t)}_1, \rho^{(t)}_2, ..., \rho^{(t)}_m\}$, according to Eq.\ref{get_arch}.
And the skipped MHAs in the $j$-th subnetwork can be represent as $\rho^{(t)}_j=\{\mu_{j,1}, \mu_{j,2}, ..., \mu_{j,m}\}$.
Based on the loss values $l^{(t)}_j$ of the $j$-th subnetwork, the reward of it is defined as $e^{-l^{(t)}_j}$, which is proportional to subnetwork performance.
According to the rewards, the action preference of MHA modules can be updated by
\begin{equation}
\begin{aligned}
s^{(t)}_i = & s^{(t-1)}_i + (\frac{1}{m}\sum_{h=1}^{m}r^{(t)}_h - r^{(t)}_j) \pi^{(t)}_{j,i} (1 - \pi^{(t)}_{j,i}), \\
& \mathrm{where} \ \ i \notin \rho^{(t)}_j, \rho^{(t)}_j \in \Phi^{(t)}.
\label{update}
\end{aligned}
\end{equation}
When the reward of the sampled subnetwork is larger than the average one $\frac{1}{m} \sum_{j=1}^m e^{-l^{(t)}_j}$, it suggests that the activated MHA modules are more important.
The change of their action preference in this step can be ontained by $\frac{1}{m} \sum_{j=1}^m e^{-l^{(t)}_j} - e^{-l^{(t)}_i}$.
To smooth optimization, we introduce $\pi^{(t)}_i (1 - \pi^{(t)}_i)$ to control the learning rate.

%
Finally, we skip the $k$ layers with the lowest action preference $\mathbf{s}$.
The search process can find out the redundancy in the original model for the downstream task
By skipping the most redundant $k$ layers, we can gain the accelerated model in the downstream task.

\begin{table*}[t]
\caption{
Comparison among EAS and other PETL methods for ScienceQA on LLaVA. 
The corner marks of PCETL methods represent the number of skipped modules.
``\emph{Inference Time}'' refers to the time per sample. 
The best and second best results are marked in \textbf{bold} and \underline{underline}, respectively.
}
\resizebox{1.0\textwidth}{!}
{
\newcommand{\bxmark}{\usym{2613}} %
\setlength{\tabcolsep}{2pt}
\begin{tabular}{l c c c |c c c|c c c|c c|c }

\toprule \multirow{2}{*}{\makecell[c]{\textbf{Method}} } & \multirow{2}{*}{\makecell[c]{\textbf{Updated}\\\textbf{Parameters}}} 
& \multirow{2}{*}{ \makecell[c]{\textbf{FLOPs}}}
& \multirow{2}{*}{ \makecell[c]{\textbf{Inference} \\ \textbf{Time}}}
& \multicolumn{3}{c|}{ \textbf{Subject} } & 
\multicolumn{3}{c|}{ \textbf{Context Modality} } & \multicolumn{2}{c|}{ \textbf{Grade} } & \multirow{2}{*}{ \textbf{Average} } \\

& & & & \textbf{NAT} & \textbf{SOC} & \textbf{LAN} & \textbf{TXT} & \textbf{IMG} & \textbf{NO} &\textbf{ G1-6 }&\textbf{ G7-12 }& \\

\midrule \multicolumn{12}{l}{\textit{Fully tuned methods}}\\
LLaVA-7B \cite{liu2023visual}           & 100.00\%    & 5018.4G {\scriptsize(+0.0G)} & 2.19s & {90.4} & {96.0} & {88.0} & 89.5 & {88.0} & 90.7 & 90.9 & {90.9} & {90.9} \\

\midrule \multicolumn{12}{l}{\textit{Parameter-efficient methods}} \\
LoRA \cite{conf/iclr/HuSWALWWC22}    & 0.62\% & 5018.4G {\scriptsize(+0.0G)} & 2.34s & 86.5 & 95.1 & \underline{85.9} & 86.1 & 86.6 & 87.0 & 88.7 & \underline{87.2} & 88.1
\\
Adapter \cite{conf/cvpr/Sung0B22} & 0.54\% & 5041.6G {\scriptsize(+23.2G)} & 2.89s & 86.9 & 95.5 & 85.8 & 86.4 & 86.3 & 87.7 & 89.5 & 86.5 & 88.4

\\
MAM \cite{conf/iclr/HeZMBN22} & 0.54\% & 5041.6G {\scriptsize(+23.2G)} & 2.87s & \underline{87.3} & \underline{95.7} & \textbf{86.6} & 86.5 & 86.1 & \textbf{88.9} & \underline{89.6} & \textbf{87.7} & \textbf{88.9}
 \\
\midrule \multicolumn{12}{l}{\textit{Parameter and computation efficient methods}} \\
DAS$_{4}$ \cite{wu2023parameter}     & 0.59\% & 4475.0G {\scriptsize(-543.4G)} & 2.53s & \textbf{87.5} & 95.5 & 84.5 & \textbf{87.4} & 87.0 & 86.4 & 89.2 & {87.0} & 88.4 \\
\rowcolor{gray!20}
EAS$_{8}$ (Ours)                     & 0.78\% & 4616.3G {\scriptsize(-402.1G)} & 1.98s & 87.2 & \textbf{96.1} & 85.6 & \underline{86.6} & \textbf{87.2} & \underline{87.4} & \textbf{89.7} & 86.8 & \underline{88.7} \\
\rowcolor{gray!20}
EAS$_{12}$ (Ours)                    & 0.90\% & 4415.4G {\scriptsize(-603.0G)} & 1.82s & 86.6 & 94.9 & 85.6 & 86.2 & 86.6 & 86.8 & 88.6 & \underline{87.2} & 88.1 \\
\bottomrule[1.00pt]
\end{tabular}
}
\label{compare_LLaVA_on_ScienceQA}
\end{table*}

\begin{table*}[t]
\caption{
Comparison among EAS and other PETL-Based MLLMs for ScienceQA. 
The corner marks of PCETL methods represent the number of skipped modules.
``\emph{Inference Time}'' refers to the time per sample. 
The best and second best results are marked in \textbf{bold} and \underline{underline}, respectively.
}
\resizebox{1.0\textwidth}{!}
{
\newcommand{\bxmark}{\usym{2613}} %
\setlength{\tabcolsep}{2pt}
\begin{tabular}{l c c c |c c c|c c c|c c|c }

\toprule \multirow{2}{*}{\makecell[c]{\textbf{Method}} } & \multirow{2}{*}{\makecell[c]{\textbf{Updated}\\\textbf{Parameters}}} 
& \multirow{2}{*}{ \makecell[c]{\textbf{FLOPs}}}
& \multirow{2}{*}{ \makecell[c]{\textbf{Inference} \\ \textbf{Time}}}
& \multicolumn{3}{c|}{ \textbf{Subject} } & 
\multicolumn{3}{c|}{ \textbf{Context Modality} } & \multicolumn{2}{c|}{ \textbf{Grade} } & \multirow{2}{*}{ \textbf{Average} } \\

& & & & \textbf{NAT} & \textbf{SOC} & \textbf{LAN} & \textbf{TXT} & \textbf{IMG} & \textbf{NO} &\textbf{ G1-6 }&\textbf{ G7-12 }& \\

\midrule \multicolumn{12}{l}{ \textit{Fully tuned methods} }\\
LLaVA-7B \cite{liu2023visual}              & 100.00\%    & 5018.4G {\scriptsize(+3303.5G)} & 2.19s & {90.4} & {96.0} & {88.0} & 89.5 & {88.0} & 90.7 & 90.9 & {90.9} & {90.9} \\

\midrule \multicolumn{12}{l}{\textit{Parameter-efficient methods}} \\
LaVIN-7B \cite{luo2023cheap}            & 0.06\% & 1715.4G {\scriptsize(+0.6G)} & 3.70s & \underline{89.3} & \underline{94.9} & 85.2 & \underline{88.5} & \underline{87.5} & 88.1 & \underline{90.2} & \underline{88.1} & \underline{89.4} \\
LLaMA-LoRA \cite{hu2022lora}            & 0.08\% & 1714.8G {\scriptsize(+0.0G)} & 2.22s & 84.1 & 78.0 & \underline{85.7}  & 82.8 & 74.1 & \underline{88.6} & 83.8 & 82.2 & 83.2 \\
LLaMA-Adapter \cite{zhang2023llama}     & 0.03\% & 1839.6G {\scriptsize(+124.7G)} & 4.35s & 84.4 & 88.3 & 84.4 & 83.7 & 80.3 & 86.9 & 85.8 & 84.1 & 85.2 \\
\midrule \multicolumn{12}{l}{\textit{ Parameter and computation efficient methods }} \\
LaVIN-DAS$_{4}$ \cite{wu2023parameter}     & 0.07\% & 1501.3G {\scriptsize(-213.6G)} & 
3.23s & 89.0 & 94.6 & 85.1 & 87.9 & 86.5 & 88.4 & 89.7 & \textbf{88.1} & 89.2 \\
\rowcolor{gray!20}
LaVIN-EAS$_{12}$ (Ours)                    & 0.11\%  & 1497.1G {\scriptsize(-217.7G)} & 1.69s & \textbf{89.5} & \textbf{95.6} & \textbf{86.4} & \textbf{88.7} & \textbf{88.3} & \textbf{88.8} & \textbf{91.3} & 87.7 & \textbf{90.0} \\
\bottomrule[1.00pt]
\end{tabular}
}
\label{compare_LaVIN_on_ScienceQA}
\end{table*}

\begin{table*}[t]
\caption{
%
Comparison among EAS and other PETL methods on Slake and AID. 
The base model used is LLaVA-7B.
%
%
``ZS'' and ``FT'' denote ``zero-shot'' and ``fine-tuned'', receptively.
The best results of efficient tuning methods are marked in \textbf{bold} and the second best is \underline{underline}, respectively. 
}
\centering
\resizebox{1.0\textwidth}{!}
{
\setlength{\tabcolsep}{2pt}
\begin{tabular}{l c | c c c | c c c | c c }
\toprule
\multirow{3}{*}{\textbf{Methods}}
& \multirow{3}{*}{\makecell[c]{\textbf{Updated} \\ \textbf{Parameter}}}
& \multicolumn{3}{c|}{\textbf{Slake}} & \multicolumn{3}{c|}{\textbf{AID}} & \multicolumn{2}{c}{\textbf{Average}} \\
& & \multirow{2}{*}{\makecell[c]{\textbf{FLOPs}}} & \multirow{2}{*}{\makecell[c]{\textbf{Accuracy} \\ \textbf{(Open)}}} & \multirow{2}{*}{\makecell[c]{\textbf{Accuracy} \\ \textbf{(Closed)}}} & \multirow{2}{*}{\makecell[c]{\textbf{FLOPs}}} & \multirow{2}{*}{\makecell[c]{\textbf{Accuracy} \\ \textbf{(20\%)}}} & \multirow{2}{*}{\makecell[c]{\textbf{Accuracy} \\ \textbf{(50\%)}}} & \multirow{2}{*}{\makecell[c]{\textbf{FLOPs}}} & \multirow{2}{*}{\makecell[c]{\textbf{Accuracy}}} \\
& & & & & & & & &\\
\midrule \multicolumn{10}{l}{\textit{ Fully tuned methods }} \\
LLaVA-7B-ZS \cite{liu2023visual}               & 100.00\%  & 4904.0G {\scriptsize(+0.0G)} & 28.7 & 56.5 & 4869.9G {\scriptsize(+0.0G)} & 25.5 & 25.6 & 4886.9G {\scriptsize(+0.0G)} & 32.6 \\
LLaVA-7B-FT \cite{liu2023visual}               & 100.00\%  & 4904.0G {\scriptsize(+0.0G)} & 80.3 & 85.6 & 4869.9G {\scriptsize(+0.0G)} & 97.7 & 97.9 & 4886.9G {\scriptsize(+0.0G)} & 90.4 \\
\midrule \multicolumn{10}{l}{\textit{ Parameter-efficient methods }} \\
LoRA \cite{conf/iclr/HuSWALWWC22}              & 0.61\%    & 4904.0G {\scriptsize(+0.0G)}   & \textbf{79.2} & 85.6 & 4869.9G {\scriptsize(+0.0G)} & 96.9 & 97.3  & 4886.9G {\scriptsize(+0.0G)} & {89.8} \\
Adapter \cite{conf/cvpr/Sung0B22}              & 0.53\%    & 4926.7G {\scriptsize(+22.7G)} & 77.5 & 87.3 & 4892.4G {\scriptsize(+22.5G)} & 96.2 & 96.8 & 4909.5G {\scriptsize(+22.6G)} & 89.5 \\
MAM \cite{conf/iclr/HeZMBN22}                  & 0.53\%    & 4926.7G {\scriptsize(+22.7G)} & 77.4 & \textbf{87.7} & 4892.4G {\scriptsize(+22.5G)} & 96.2 & 97.7 & 4909.5G {\scriptsize(+22.6G)} & {89.8} \\
\midrule \multicolumn{10}{l}{\textit{ Parameter and computation efficient methods }} \\

DAS$_4$ \cite{wu2023parameter}              & 0.63\%    & 4373.0G {\scriptsize(-531.0G)}  & \textbf{79.2} & 86.5 & 4342.9G {\scriptsize(-527.0G)} & 96.6 & \underline{97.8} & 4358.0G {\scriptsize(-529.0G)} & \underline{90.0} \\
\rowcolor{gray!20}
EAS$_{10}$ (Ours)                             & 0.85\%  & 4412.9G {\scriptsize(-491.1G)} & 78.1 & 87.0 & 4406.9G {\scriptsize(-497.1G)} & \textbf{97.5} & \textbf{98.0} & 4409.9G {\scriptsize(-477.1G)} & \textbf{90.2} \\
\rowcolor{gray!20}
EAS$_{12}$ (Ours)                             & 0.86\%  & 4314.7G {\scriptsize(-589.3G)} & 77.4 & \underline{87.5} & 4308.8G {\scriptsize(-561.1G)} & \underline{97.5} & 97.58 & 4311.8G {\scriptsize(-575.2G)} & 89.5 \\
\bottomrule[1.00pt]\
\end{tabular}
}
\label{table_llava}
\end{table*}

\begin{table*}[t]
\caption{
Comparison among EAS, DAS and PETL methods for METER on VQA, NLVR$^2$ and Flickr30K.
The best and second performance is \textbf{bold} and \underline{underlined}.
}
\centering
\setlength{\tabcolsep}{2pt}
\resizebox{1.0\textwidth}{!}
{
\begin{tabular}{l  c | cc cc cc | cc}
\toprule
\multirow{2}{*}{\textbf{Method}} 
& \multirow{2}{*}{\makecell[c]{\textbf{Updated} \\ \textbf{Parameters}}} 
& \multicolumn{2}{c}{\textbf{VQA}} 
& \multicolumn{2}{c}{\textbf{NLVR$^2$}} 
& \multicolumn{2}{c|}{\textbf{Flickr30K}} 
& \multicolumn{2}{c}{\textbf{Average}} \\
& & \makecell[c]{\textbf{FLOPs}} & \textbf{test-dev} & \makecell[c]{\textbf{FLOPs}} & \textbf{test-P} & \makecell[c]{\textbf{FLOPs}} & \textbf{IR/TR R@1} & \makecell[c]{\textbf{FLOPs}} & \textbf{Accuracy} \\
\midrule \multicolumn{10}{l}{\textit{ Full tuned methods }} \\
Full Tuning                                     & 100.00\%    & 93.2G {\scriptsize(+0.0G)}   & 77.4  & 52.9G {\scriptsize(+0.0G)}  & 83.1 & 93.2G {\scriptsize(+0.0G)} & 82.2/94.3 {\scriptsize(+0.0G)}  & 79.8G {\scriptsize(+0.0G)}  & 84.3  \\
\midrule \multicolumn{10}{l}{\textit{ Parameter-efficient methods }} \\
Shallow Prompt~\cite{conf/acl/LiL20}            & 0.09\%   & 121.9G {\scriptsize(+28.7G)} & 68.5  & 79.7G {\scriptsize(+26.9G)} & 65.7 & 121.9G {\scriptsize(+28.7G)}  & 74.2/88.6 & 107.9G {\scriptsize(+28.7G)} & 74.3 \\
Deep Prompt~\cite{journals/corr/abs-2203-12119} & 0.57\%   & 99.8G {\scriptsize(+6.5G)}   & 70.8  & 58.5G {\scriptsize(+5.6G)}  & 72.6 & 99.8G {\scriptsize(+6.5G)}    & 78.8/89.4 & 86.0G {\scriptsize(+6.2G)} & 77.9 \\
LoRA~\cite{conf/iclr/HuSWALWWC22}               & 0.09\%   & 93.2G {\scriptsize(+0.0G)}   & 74.0  & 52.9G {\scriptsize(+0.0G)}  & 78.8 & 93.2G {\scriptsize(+0.0G)}    & 79.9/\underline{92.6} & 79.8G {\scriptsize(+0.0G)} & 81.3 \\
Adapter~\cite{conf/cvpr/Sung0B22}               & 1.65\%   & 94.9G {\scriptsize(+1.6G)}   & 74.7  & 54.3G {\scriptsize(+1.4G)}  & 79.9 & 94.9G {\scriptsize(+1.6G)}    & \underline{80.4}/91.9 & 81.3G {\scriptsize(+1.6G)} & 81.7 \\
MAM~\cite{conf/iclr/HeZMBN22}                   & 1.11\%   & 94.3G {\scriptsize(+1.1G)}   & \textbf{75.1}  & 53.5G {\scriptsize(+0.7G)}  & \textbf{80.4} & 94.3G {\scriptsize(+1.1G)}    & \textbf{80.4}/\textbf{93.2} & 80.7G {\scriptsize(+1.0G)} & \textbf{82.3} \\
\midrule \multicolumn{10}{l}{\textit{ Parameter and computation efficient methods }} \\
DAS$_4$ \cite{wu2023parameter}                  & 1.65\%   & \underline{82.1G {\scriptsize(-11.2G)}}  & 74.8  & 47.7G {\scriptsize(-5.1G)}  & 80.1 & \underline{82.1G {\scriptsize(-11.6G)}}   & 80.1/91.8 & \underline{70.6G {\scriptsize(-9.2G)}} & 81.7 \\
\rowcolor{gray!20}
EAS$_{8}$ (Ours)                                & 1.26\%   & 82.6G {\scriptsize(-10.7G)}  & \underline{74.9}  & \underline{45.5G {\scriptsize(-7.4G)}}  & \underline{80.1} & 84.1G {\scriptsize(-9.1G)}    & \underline{80.4}/92.2 & 70.7G {\scriptsize(-9.1G)} & \underline{81.9} \\ 
\rowcolor{gray!20}
EAS$_{10}$ (Ours)                               & 1.26\%   & \textbf{78.7G {\scriptsize(-14.5G)}}  & 74.8  & \textbf{42.7G {\scriptsize(-10.2G)}} & 80.1 & \textbf{81.3G {\scriptsize(-11.9G)}}   & 79.8/90.4 & \textbf{67.6G {\scriptsize(-12.2G)}} & 81.3 \\
\bottomrule[1.00pt]
\end{tabular}
}
\label{table_METER}
\end{table*}

\section{Experiment}

\subsection{Datasets and Metrics}

\textbf{Out-domain VL Benchmarks.}
For visual-language (VL) tasks, we first apply EAS to LLaVA on two out-domain datasets: \textbf{Slake} \cite{conf/isbi/LiuZXMYW21}, which is focused on medical question answering, and \textbf{AID} \cite{journals/tgrs/XiaHHSBZZL17} which involves remote sensing image classification.
Slake \cite{conf/isbi/LiuZXMYW21} is designed for medical visual question answering, collecting images of CT, MRI, and X-ray scans across ten different parts of the human body, such as the lungs, abdomen, and brain. 
The dataset includes ten types of questions related to organ identification, location, and abnormality. 
It is split into \emph{train}, \emph{val}, and \emph{test} sets, containing $9,849$, $2,109$, and $2,070$ examples, respectively.
Depending on the nature of the answer, questions can be categorized as ``open'' or ``closed''. 
\emph{open} questions have answers in the form of words or phrases, while \emph{closed} questions have binary answers (\emph{i.e.}, yes or no).
AID \cite{journals/tgrs/XiaHHSBZZL17} is a dataset used for remote sensing images classification.
It contains 10,000 images across 30 different scene types. 
These images are collected from Google Earth and cover diverse regions worldwide. 
The dataset can be divided into subsets with ``20\%'' and ``50\%'' of samples used for training.

\noindent \textbf{Common VL Benchmarks.} 
We also evaluated EAS on several widely recognized VL benchmarks, including ScienceQA \cite{lu2022learn}, VQA2.0 \cite{conf/cvpr/GoyalKSBP17}, NLVR$^2$  \cite{conf/acl/SuhrZZZBA19}, and Flickr30K \cite{journals/ijcv/PlummerWCCHL17}.
ScienceQA \cite{lu2022learn} is a large-scale benchmark for science question answering, encompassing three main subjects, 26 topics, 127 categories, and 379 distinct skills. 
The dataset includes examples with text only as well as those with both text and images. It is split into \emph{train}, \emph{val}, and \emph{test} sets, containing 12,726, 4,241, and 4,241 examples, respectively.
Different from traditional question-answering datasets, ScienceQA requires the model to give the reasoning process and have a longer output sequence.
VQA2.0~\cite{conf/cvpr/GoyalKSBP17} is a visual question answering dataset that consists of 204,721 images, with each image linked to at least three open-ended questions.
NLVR$^2$~\cite{conf/acl/SuhrZZZBA19} is a benchmark designed to determine whether a natural language statement is true given a pair of images, effectively framing it as a binary classification task based on joint representations of the two images.
Flickr30K~\cite{journals/ijcv/PlummerWCCHL17}, re-split by Karpathy \emph{et al.}~\cite{journals/pami/KarpathyF17}, consists of 31,000 images, each paired with five descriptive sentences. This dataset is commonly used for evaluating cross-modal retrieval tasks, where the goal is to retrieve relevant text given an image or vice versa.

\subsection{Implementation Details}

The base models employed in our study are LLaVA \cite{liu2023visual} and LaVIN \cite{luo2023cheap}.
When applying EAS to LLaVA, we insert PIAs before each attention module, and set the hidden dimension of PIA to 256.
We also employ PIAs for parameter-efficient adaptations, similar to \cite{luo2023cheap, zhang2023llama}. 
These PIAs also have a hidden dimension of 128, and only one path for adaptation is used, \emph{i.e.}, $f_{d1}$ in Eq.\ref{first_half_PIA}. 
As for LaVIN, following its default settings, we use LLaMA-7B \cite{touvron2023llama} as the LLM, and adopt ViT-L \cite{conf/iclr/DosovitskiyB0WZ21} as the image encoder.
Regarding PIA, we set the hidden dimension to 32 for skipping and 8 for adaptation. 
We insert these PIAs before each attention module.
For redundancy evaluation, we first train the subnetworks with randomly skipped modules for 5 epochs.
Then we introduce redundancy evaluation in the next 5 epochs, where 3 subnetworks are sampled every 10 training steps. 
In addition, we also validate the proposed EAS on a classical VL pre-trained model named METER \cite{conf/cvpr/DouXGWWWZZYP0022}.
Its attention modules in a co-attention layer are considered as independent potential skipped components. 
The PIAs for skipping attention modules are configured with a hidden dimension of 144. 
The remaining settings following the defaults in DAS \cite{wu2023parameter}.

\subsection{Experimental Results}

\subsubsection{Quantitative Analysis}

\noindent \textbf{Comparison with PETL and PCETL methods.} 
We first compare our EAS with PCETL and PETL methods, such as DAS \cite{wu2023parameter} and Adapter \cite{conf/cvpr/Sung0B22}, on ScienceQA \cite{lu2022learn} which requires long output sequences, of which results are given in Tab.\ref{compare_LLaVA_on_ScienceQA} and Tab.\ref{compare_LaVIN_on_ScienceQA}.
We can first observe that PETL methods achieve great performance, while retaining a very small number of parameters to update.
However, the use of adapters also obviously slows down their inference speeds.
For instance, the inference of LLaMA-Adapter in Tab.\ref{compare_LaVIN_on_ScienceQA} consumes more than $4.35$ seconds for each example.
Compared with these approaches, both DAS and EAS can achieve competitive performance, while retaining parameter and computation efficiency. 
Speciffically, in Tab.~\ref{compare_LLaVA_on_ScienceQA}, DAS can skip $4$ Transformer layers of LLaVA for better efficiency compare to PETL method MAM \cite{conf/iclr/HeZMBN22} while retraining comparable performance, \emph{i.e} $-0.6\%$ accuracy while $+11.8\%$ efficiency.
Meanwhile, EAS-7B$_{12}$ can improve the inference speed by $+36.6\%$ with only $-1.0\%$ accuracy drop.

In Tab.\ref{compare_LaVIN_on_ScienceQA}, we examine EAS in multimodal transfer learning of LLMs and compare it with an PETL-based MLLM called LaVIN \cite{luo2023cheap}. 
In practice, we replace the adapter-designs in LaVIN with DAS and our EAS.  
Compared to LaVIN, DAS$_4$ significantly improves the inference efficiency by $1.14$ times with only $-0.2\%$ accuracy drop.
Furthermore, EAS$_{12}$ can achieve simialr performance while improving the inference speeds by $2.18$ times compared to LaVIN.
As can be seen, also as a PECTL method, our EAS is much superior in actual speed-up of MLLMs than DAS, while its performance retains comparable. 
We can observe that EAS is obviously superior in actual speed-up, while retaining similar performance.
When we combine the experiment results in Tab.\ref{compare_LLaVA_on_ScienceQA} and Tab.\ref{compare_LaVIN_on_ScienceQA}, we can find that the difficulty of transferring is related to the base model itself. 
Specifically, DAS$_4$ has an average performance of 88.1\% on LLaVA for ScienceQA, while the performance in LaVIN is 89.2\%, achieving a 0.4\% improvement. 
For the proposed EAS method, it can skip 12 layers on LaVIN without performance drop, but can only skip 8 layers on LLaVA.

In Tab.\ref{table_llava}, we examine EAS on two out-domain benchmarks, \emph{i.e.}, Slake \cite{conf/isbi/LiuZXMYW21} and AID \cite{journals/tgrs/XiaHHSBZZL17}, and compare it with a set of PCETL and PETL methods. 
From this table, we can first observe that the generalization of MLLMs like LLaVA is very poor on this two benchmarks, which are about 42.6\% and 25.5\%, respectively. 
This result confirms our motivation that MLLMs have a greater demand in task tuning. 
It can be seen that either using full tuning or PETL methods, the performance can be substantially improved.
Among the compared methods, the PECTL methods, \emph{i.e.}, DAS and our EAS, are the ones that can reduce computation overhead while achieving high transfer learning performance. 
In particular, the merits of our EAS are also more prominent than DAS, which can skip up to 8 MHAs with almost no performance drops to full tuning, \emph{i.e.,} 90.2 \emph{v.s.} 90.4. 
Another observation is that the impact of EAS is also influenced by the downstream tasks.
For instance, EAS can skip 12 layers without performance drop of LLaVA on Slake dataset, while only 10 layers can be skipped for AID dataset.

\begin{table}[t]
\caption{
Comparison of different redundancy evaluation candidates for EAS on ScienceQA.
``Num. of Skip'' refers to the number of skipped modules.
``Updated Param.'' refers to the scale of updated parameters.
``Infer. Time'' refers to the inference time.
}
\setlength{\tabcolsep}{3pt}
\resizebox{0.95\columnwidth}{!}{
\begin{tabular}{c | c c c c}
\toprule[1.00pt]
\makecell[c]{\textbf{Candidate} \\ \textbf{Modules}} 
& \makecell[c]{\textbf{Num. of}\\ \textbf{Skip} } 
& \makecell[c]{\textbf{Updated} \\ \textbf{Param.}} 
& \makecell[c]{\textbf{Infer.} \\ \textbf{Time}} 
& \makecell[c]{\textbf{Average} \\ \textbf{Performance}} \\
\midrule
\multirow{3}{*}{MHA}        & 4     & 0.08\% & 2.08s & 89.4 \\
                            & 8     & 0.10\% & 1.91s & 89.8 \\
                            & 12    & 0.11\% & 1.69s & 90.0 \\
\midrule                
\multirow{3}{*}{FFN}        & 4     & 0.08\% & 2.26s & 88.3 \\
                            & 8     & 0.11\% & 2.18s & 85.0 \\
                            & 12    & 0.13\% & 2.11s & 82.1 \\
\midrule
\multirow{3}{*}{MHA or FFN} & 4     & 0.08\% & 2.13s & 88.4 \\
                            & 8     & 0.10\% & 2.09s & 89.2 \\
                            & 12    & 0.12\% & 1.78s & 88.6 \\
\bottomrule[1.00pt]
\end{tabular}
\label{candidate_skipped}
}
\end{table}

\begin{figure}[t]
\centering
\includegraphics[width=1.05\columnwidth]{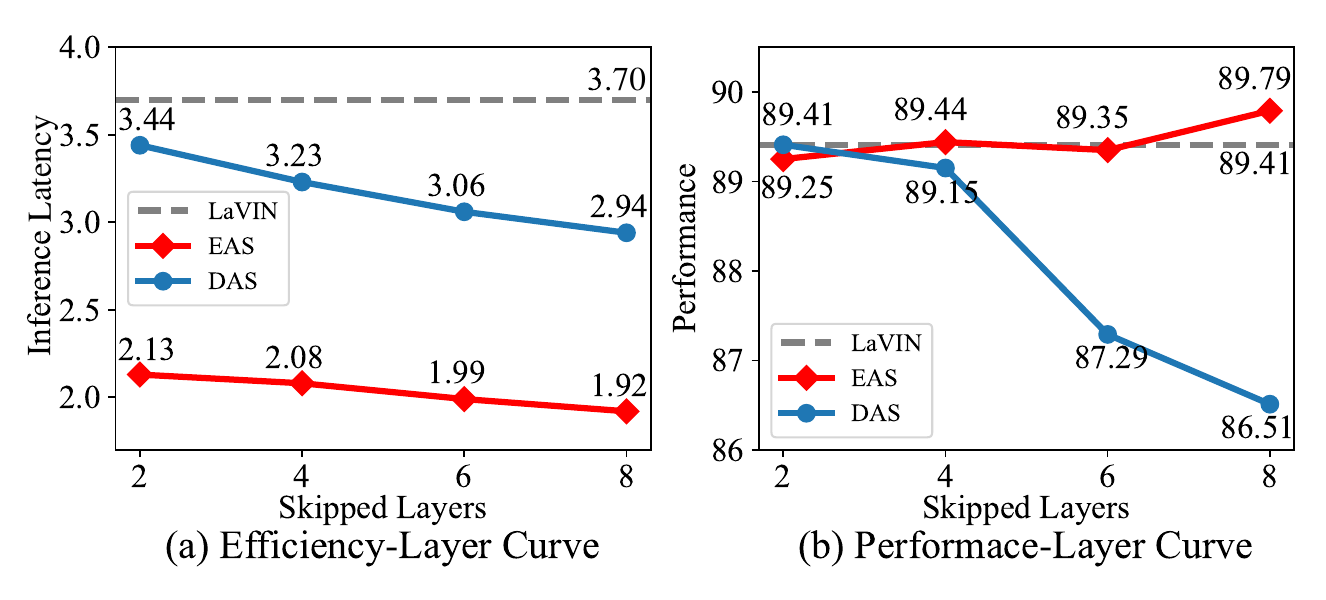}
\captionof{figure}{
\small
Comparison of the inference time and performance between EAS and DAS on ScienceQA dataset.
EAS can better speed up the inference while maintaining the performance.
}
\label{compare_to_DAS}
\vspace{-3mm}
\end{figure}

Akin to DAS \cite{wu2023parameter}, we also generalize our EAS to a classical and smaller VL pretrained model called METER \cite{conf/cvpr/DouXGWWWZZYP0022} on three widely-used VL benchmarks in Tab. \ref{table_METER}. 
Compared with the popular PETL methods \cite{journals/corr/abs-2203-12119, conf/iclr/HuSWALWWC22, conf/iclr/HeZMBN22}, our EAS can also maintain a small scale of updated parameters, while achieving comparable performance but better computation efficiency. 
For instance, compared with adapter, our EAS is slightly better in performance, \emph{i.e.}, 81.9 \emph{v.s.} 81.7, but EAS can save up to 13.0\% FLOPs. 
In terms of DAS, the advantages of our EAS are also obvious. 
For instance, retaining similar performance, EAS$_{8}$ and EAS$_{12}$ can reduce more computation by 8.0\% and 12.0\%, respectively.  
Overall, these experimental results well validate the generalization of EAS on common VL models.

\noindent \textbf{Comparisons of Actual Inference Speeds.} 
In Fig.~\ref{compare_to_DAS}, we further compare the inference speeds of our EAS and DAS on LaVIN with different settings.
The first observation from Fig.~\ref{compare_to_DAS} is that DAS can skip certain layers of LaVIN with limited performance drops, \emph{e.g.}, skipping 2-4 layers. 
Its efficiency gains become more obvious when skipping more than 6 Transformer layers, while the performance also decreases to a certain extent.
Compared with DAS, EAS has obvious merits in both performance and efficiency. 
For instance, with the same number of skipped modules, EAS is consistently faster than DAS by $+44.4\%$ to $51.6\%$. 
By skipping 8 MHAs, EAS can obtain more obvious performance gains than the default LaVIN, \emph{i.e.}, 89.8 \emph{v.s.} 89.4, while the inference speed is about $1.92$ times faster. 
Overall, these results well validate the motivation and designs of EAS towards PCETL of MLLMs.

\begin{table}[t]
\captionof{table}{
Ablation of \emph{Propagate-Information Adapter} (PIA) on EAS-7B$_{12}$. Here, \emph{Base} refers  to using only $f_{d1}$ in Eq.~\ref{first_half_PIA}. 
``\emph{Num. of Skip}'' refers to the number of skipped modules.
``\emph{Infer. Time}'' refers to the inference time.
}
\resizebox{0.95\columnwidth}{!}{
\setlength{\tabcolsep}{2pt}
\begin{tabular}{l | c c c}
\toprule
\makecell[c]{\textbf{Setting}} 
& \makecell[c]{\textbf{Updated} \\ \textbf{Param.}} 
& \makecell[c]{\textbf{Infer.} \\ \textbf{Time}} 
& \makecell[c]{\textbf{Average} \\ \textbf{Performace}} \\
\midrule
Base                                                  & 0.09\% & 2.94s & 87.1 \\
\ + Info. Exc. in Eq.\ref{first_half_PIA} \ & 0.11\% & 3.23s & 90.1 \\
\ + Frozen bias in Eq.\ref{reparameter_down} \   & 0.11\% & 2.97s & 90.0 \\
\ + Re-param.                                & 0.11\% & 1.69s & 90.0 \\
\bottomrule
\end{tabular}
\label{ablation_on_component}
}
\end{table}

\begin{table}[t]
\captionof{table}{
Ablation of the skipping number on ScienceQA.
``Updated Param.'' refers to the scale of updated parameters.
``Infer. Time'' refers to the inference time.
}
\resizebox{0.95\columnwidth}{!}{
\setlength{\tabcolsep}{2.2mm}
\begin{tabular}{l | c c c}
\toprule
\textbf{Methods} 
& \makecell[c]{\textbf{Updated} \\ \textbf{Param.}} 
& \makecell[c]{\textbf{Infer.} \\ \textbf{Time}} 
& \makecell[c]{\textbf{Average} \\ \textbf{Performance}} \\
\midrule
LaVIN-7B        & 0.06\% & 3.70s & 89.4 \\
\midrule
EAS-7B$_{4}$    & 0.08\% & 2.08s & 89.4 \\
EAS-7B$_{8}$    & 0.10\% & 1.92s & 89.8 \\
EAS-7B$_{12}$   & 0.11\% & 1.69s & 90.0 \\
EAS-7B$_{16}$   & 0.13\% & 1.56s & 88.4 \\
\bottomrule
\end{tabular}
\label{ablation_on_skip_num}
}
\end{table}

\noindent \textbf{Ablation Study.} 
To examine the designs of EAS, we further conduct extensive ablations in Tab.\ref{candidate_skipped}-\ref{ablation_on_skip_num}. 
In Tab.\ref{candidate_skipped}, we first ablate the choice of evaluation candidates for EAS on ScienceQA, including ``\emph{MHA}'', ``\emph{FFN}'' and ``\emph{MHA or FFN}''. 
From this table, we can first see that skipping MHA is the best choice among three candidates.
Notably, EAS can skip up to 12 MHAs with even better performance, while its efficiency is also much better than the others. 
In terms of FFN, its removal leads to slower inference speed than that of MHA.
More importantly, when dropping more than 8 FFNs, the performance declines significantly, \emph{e.g.}, $-6.99\%$ by skipping 12 FFNs. 
This result well confirms our argument about the roles of MHA and FFN in MLLMs.
In terms of ``\emph{MHA or FFN}'', this is a suboptimal candidate for EAS, which can obtain a good trade-off between performance and efficiency, but it is still worse than ``\emph{MHA}''. 
Overall, these results well validate the motivation of EAS, \emph{i.e.}, not all attention is needed for MLLMs. 

\begin{table}[t]
\caption{
The impact of different hidden dimensions in PIA for skip-connections of EAS on ScienceQA. 
``Hidden Dim.'' refers to the number of hidden dimension.
``Updated Param.'' refers to the scale of updated parameters.
``Infer. Time'' refers to the inference time.
}

\resizebox{\columnwidth}{!}{
\setlength{\tabcolsep}{1mm}
\begin{tabular}{l | c c c c}
\toprule
\textbf{Methods} 
& \makecell[c]{\textbf{Hidden} \\ \textbf{Dim.}}
& \makecell[c]{\textbf{Updated} \\ \textbf{Param.}} 
& \makecell[c]{\textbf{Infer.} \\ \textbf{Time}} 
& \makecell[c]{\textbf{Average} \\ \textbf{Performance}} \\
\midrule
\multirow{4}{*}{ EAS-7B$_{8}$}  & 8  & 0.04\% & 1.89s & 87.6 \\
                                & 16 & 0.05\% & 1.91s & 88.0 \\
                                & 32 & 0.08\% & 1.92s & 89.8 \\
                                & 64 & 0.22\% & 1.94s & 90.1 \\
\bottomrule
\end{tabular}
}
\label{ablation_on_hidden_dimension}
\end{table}

\begin{figure}[t]
\centering
\includegraphics[width=1.0\columnwidth]{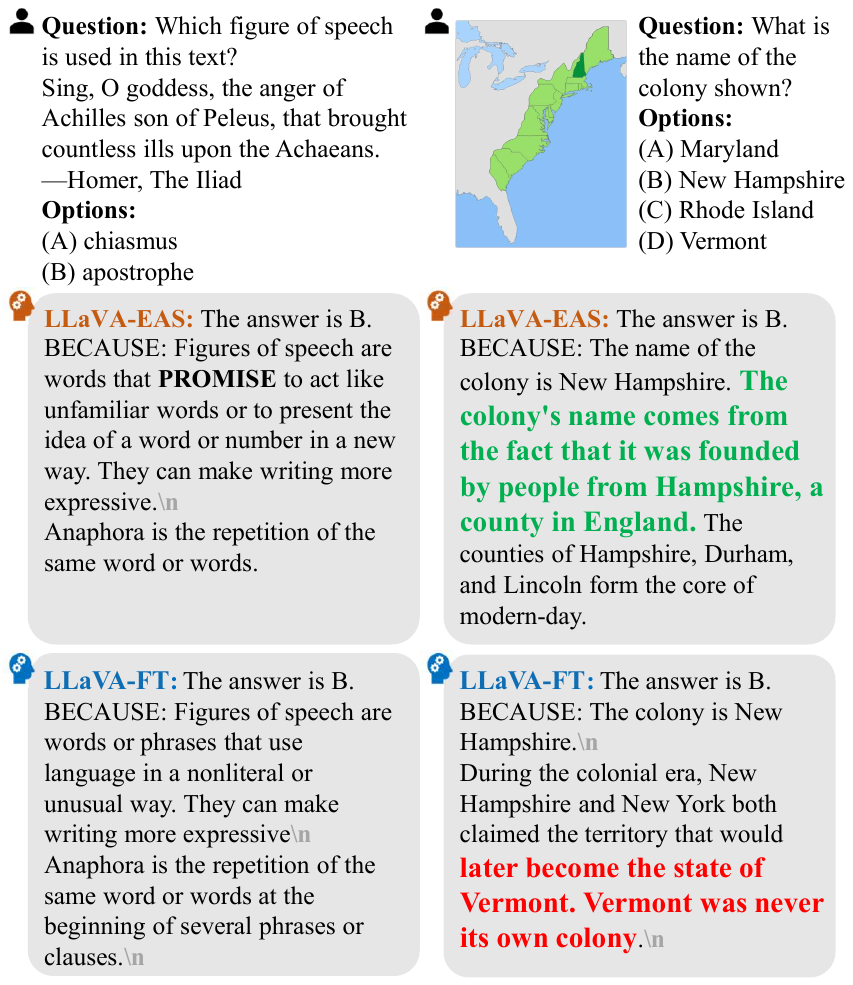}
\caption{
The predictions of LLaVA-7B-EAS and LLaVA-7B-FT.
The suffixes denotes the training by our EAS and full tuning (FT), respectively. 
The accurate explanations for the answer are highlight in \textcolor[rgb]{0.0,0.69,0.31}{green}, while the logically incorrect ones in \textcolor{red}{red}. 
}
\label{fig4:QAexample}
\end{figure}

Tab.~\ref{ablation_on_component} ablates the designs of PIA, which is the key component of EAS. 
Here, ``\emph{Base}'' refers to using only one path in the downsampling, \emph{i.e.}, only $f_{d1}$ in Eq.~\ref{first_half_PIA}.
``\emph{Info. Exc.}'' refers to the information exchange in PIA, \emph{i.e.}, Eq.~\ref{first_half_PIA}.
``\emph{Frozen bias}'' refers to fixing the output values of average pooling in Eq.~\ref{reparameter_down} after the first forward.
``\emph{Re-param}'' refers to applying the re-parameter in inference.
We can see that without the two-path design for information exchange (\emph{Info. Exc.}), its performance declines greatly, \emph{i.e.}, $-3.30\%$.
This phenomenon shows that adding global pooling-based information propagation operations in PIA can effectively alleviate the damage caused by over-simplifying the attention mechanism.
In PIA, a key step for its re-parameterization is freeze the dynamic bias term in Eq.\ref{reparameter_down}, \emph{i.e.}, \emph{``Frozen bias''}, so that PIA can be re-parameterized into the closest linear layer in the first forward.
As shown in Tab.\ref{ablation_on_component}, this operation has very limited impact on performance, \emph{i.e.}, only -0.1. 
Thus, the following re-parameterization (\emph{re-param}) can be further executed to achieve actual speed-up with the same performance.
Overall, Tab.~\ref{ablation_on_component} well confirms the designs of information exchange and re-parameterization in PIA.  

In Tab.~\ref{ablation_on_skip_num}, we report the results of skipping different
numbers of MHAs by EAS.
The first observation is that skipping appropriate MHA modules has little impact on performance, \emph{e.g.} skipping up to 16 MHAs only has about $1.1\%$ performance drops, strongly suggesting that PIA is a good substitute for MHA.
Notably, skipping a certain number of MHAs can achieve even better performance than the default LaVIN, \emph{e.g.} -12 layers with $+2.64$ accuracy.
It might be due to the great redundancy of MLLMs on ScienceQA, which also suggests that PIA can help MLLMs learn better patterns from VL data.
We can also observe that EAS can improve efficiency without performance drop within a certain number of skipped layers. 
Specifically, EAS can skip less than 12 MHAs in LaVIN to accelerate inference latency without performance drop.
Again, these results confirm the effectiveness of EAS towards PCETL.

Tab. \ref{ablation_on_hidden_dimension} reports the impact of different trainable parameter sizes on EAS via setting the hidden dimension. 
From the table, we can observe that more performance gains can be obtained via using more updated parameters, \emph{e.g.} the performance is improved by $+2.64$ with $0.18\%$ updated parameters more.
This result also suggests that replacing redundant MHAs still need a certain model capacity to accommodate. 
Even so, the optimal solution, \emph{i.e.}, skipping 12 MHAs, still consumes a very small proportion of parameters, \emph{i.e.}, $0.11\%$, showing high parameter efficiency of EAS.

\subsubsection{Qualitative Analysis}

In Fig.~\ref{fig4:QAexample}, we visualize the language-only and image-language examples of LLaVA-7B with and without our EAS$_{12}$.
%
From these examples, we can see that via either full tuning or EAS tuning, LLaVA-7B correctly answer the question. 
Meanwhile, these two settings all keep the strong language ability of LLaMA \cite{touvron2023llama}, and can explain the answers fluently. 
Notably, for the visual question (right), LLaVA-EAS has better and more detailed explanations for the answer, \emph{e.g.}, ``\emph{The colony's name comes from the fact that it was founded by people from Hampshire, a county in England}''.  
In contrast, the response of LLaVA-FT is logically incorrect even through its answer is right. 
Overall, these results confirm again the merits of EAS for the efficient adaption of MLLMs.

\section{Conclusion} 

In this paper, we propose a novel method for parameter and computation efficient tuning of \emph{Multi-modal Large Language Models} (MLLMs), named by \emph{Effective Attention Skipping}.
We first reveal that not all MHAs are necessary for the efficient adaption of MLLMs, based on which EAS adopts a granular redundancy evaluation scheme. 
Meanwhile, to avoid the additional computation caused by the adapter-based skip connections, EAS is also equipped with a novel \emph{Propagation-of-Information Adapter} (PIA), which can not only keep parameter efficiency but also can be re-parameterized into the model without extra latency. 
To validate EAS, we apply it to recent MLLMs including LLaVA and LaVIN, and conduct extensive experiments on a set of VL benchmarks and two out-domain datasets. 
The experimental results show that EAS can achieve better performance than LaVIN while speeding up inference by up to 2.18 times, and EAS also improves the inference speed by 1.20 times while maintaining performance compared to LLaVA.  

\section{Acknowledgments}

This work was supported by the National Science Fund for Distinguished Young Scholars (No.62025603), the National Natural Science Foundation of China (No. U21B2037, No. U22B2051, No. U23A20383, No. U21A20472, No. 62176222, No. 62176223, No. 62176226, No. 62072386, No. 62072387, No. 62072389, No. 62002305 and No. 62272401), the Natural Science Foundation of Fujian Province of China (No. 2021J06003, No.2022J06001) and the Fundamental Research Funds for the Central Universities (Xiamen University: No. 20720240053).









\section{Declarations}


\begin{itemize}
\item Funding: This work was supported by the National Science Fund for Distinguished Young Scholars (No.62025603), the National Natural Science Foundation of China (No. U21B2037, No. U22B2051, No. U23A20383, No. U21A20472, No. 62176222, No. 62176223, No. 62176226, No. 62072386, No. 62072387, No. 62072389, No. 62002305 and No. 62272401), the Natural Science Foundation of Fujian Province of China (No. 2021J06003, No.2022J06001) and the Fundamental Research Funds for the Central Universities (Xiamen University: No. 20720240053).
\item Conflict of interest/Competing interests: The authors have no relevant financial or non- financial interests to disclose.
\item Ethics approval and consent to participate: The authors have no relevant ethics approval to disclose.
\item Consent for participate: All authors agreed to participate in this work and made clear contributions.
\item Consent for publication: All authors agreed with the content and that all gave explicit consent to submit and that they obtained consent from the responsible authorities at the institute/organization where the work has been carried out.
\item Data availability: The datasets analysed during the current study are available in these repositories:

ScienceQA \cite{lu2022learn}
\url{https://github.com/lupantech/ScienceQA};

Slake \cite{conf/isbi/LiuZXMYW21} \url{https://huggingface.co/datasets/BoKelvin/SLAKE};

AID \cite{journals/tgrs/XiaHHSBZZL17}
\url{https://huggingface.co/datasets/blanchon/AID};

VQAv2 \cite{conf/cvpr/GoyalKSBP17}
\url{https://visualqa.org/}

NLVR$^2$ \cite{conf/acl/SuhrZZZBA19}
\url{https://lil.nlp.cornell.edu/nlvr/}

Flickr30K \cite{journals/ijcv/PlummerWCCHL17}
\url{https://bryanplummer.com/Flickr30kEntities/}

\item Code availability: Our code is publicly released at 
\url{https://github.com/DoubtedSteam/EAS}.
\item Author contribution: All authors contributed to the study's conception and design. Qiong Wu, Weihao Ye and Yiyi Zhou performed the experiment, data collection and analysis. 
Qiong Wu and Weihao Ye wrote the first draft of the manuscript, and all authors commented on previous versions of the manuscript. 
All authors read and approved the final manuscript. More detailed contributions of each author are listed below: 
Conceptualization: Qiong Wu, Yiyi Zhou and Xiaoshuai Sun; 
Methodology: Qiong Wu, Weihao Ye and Yiyi Zhou;
Writing-original draft preparation: Qiong Wu and Weihao Ye;
Writing-review and editing: Yiyi Zhou, Xiaoshuai Sun and Rongrong Ji;
Supervision: Rongrong Ji.
\end{itemize}

\bibliography{sn-bibliography}

\end{document}